\documentclass[twocolumn, twocolappendix]{aastex631}

\DeclareRobustCommand{\ion}[2]{\textup{#1\,\textsc{\lowercase{#2}}}}
\newcommand*\element[1][]{%
  \def\aa@element@tr{#1}%
  \aa@element
}
\usepackage{CJK}
\usepackage{amsmath}
\usepackage{booktabs} 
\usepackage{tablefootnote}
\usepackage{lipsum}
\usepackage{threeparttable}

\begin{document}
\begin{CJK*}{UTF8}{gbsn}
\defcitealias{kolwa2023}{K23}

\title{Gas-Poor Hosts and Gas-Rich Companions of $z\approx$3.5 Radio Active Galactic Nuclei: \\ ALMA Insights into Jet Triggering and Feedback }

\correspondingauthor{Wuji Wang}
\email{wujiwang@ipac.caltech.edu}

\author[0000-0002-7964-6749]{Wuji Wang (王无忌)}
\affiliation{IPAC, M/C 314-6, California Institute of Technology, 1200 East California Boulevard, Pasadena, CA 91125, USA}

\author[0000-0002-6637-3315]{Carlos De Breuck}
\affiliation{European Southern Observatory, Karl Schwarzschild strasse 2, 85748 Garching, Germany}

\author[0000-0003-2212-6045]{Dominika Wylezalek}
\affiliation{Zentrum f\"ur Astronomie der Universit\"at Heidelberg, Astronomisches Rechen-Institut, M\"onchhofstr 12-14, D-69120 Heidelberg, Germany}

\author[0000-0003-1939-5885]{Matthew D. Lehnert}
\affiliation{Université Lyon 1, ENS de Lyon, CNRS UMR5574, Centre de Recherche Astrophysique de Lyon, F-69230 Saint-Genis-Laval, France}

\author[0000-0002-9382-9832]{Andreas L. Faisst}
\affiliation{IPAC, M/C 314-6, California Institute of Technology, 1200 East California Boulevard, Pasadena, CA 91125, USA}

\author[0000-0002-0710-3729]{Andrey Vayner}
\affiliation{IPAC, M/C 314-6, California Institute of Technology, 1200 East California Boulevard, Pasadena, CA 91125, USA}

\author[0000-0001-5783-6544]{Nicole Nesvadba}
\affiliation{Universit\'{e} de la C\^{o}te d’Azur, Observatoire de la C\^{o}te d’Azur, CNRS, Laboratoire Lagrange, Bd de l’Observatoire, CS 34229, F-06304 Nice cedex 4, France}

\author[0000-0002-8639-8560]{Jo\"el Vernet}
\affiliation{European Southern Observatory, Karl Schwarzschild strasse 2, 85748 Garching, Germany}

\author[0000-0001-6020-4417]{Pranav Kukreti}
\affiliation{Zentrum f\"ur Astronomie der Universit\"at Heidelberg, Astronomisches Rechen-Institut, M\"onchhofstr 12-14, D-69120 Heidelberg, Germany}

\author[0000-0003-2686-9241]{Daniel Stern}
\affiliation{Jet Propulsion Laboratory, California Institute of Technology, 4800 Oak Grove Drive, Pasadena, CA 91109, USA}



\begin{abstract}
Cold gaseous systems play important roles in galaxy evolution by possibly providing fuel to ignite active galactic nuclei (AGN) activity and star-formation. In this work, we analyze [\ion{C}{ii}]$158\rm \mu m$ and continuum observations from ALMA for a sample of four radio AGN at $z \approx 3.5$, focusing on eight associated companion cloud systems discovered within projected distances of tens of kiloparsecs or less. The spatial distribution of these companions indicates that the majority of cold gas is not located at the AGN position, i.e., not in their host galaxies. With the assistance of [\ion{C}{ii}] at $0.2\arcsec$ resolution, we further confirm the gas-poor nature of the hosts by re-analyzing archival [\ion{C}{i}] (a tracer of H$_{2}$) at $\sim2\arcsec$ resolution. Our sample has [\ion{C}{ii}] luminosities in a range of $2.8\times10^{8}<L_{[\ion{C}{ii}]}/L_{\odot}<4.2\times10^{9}$. The $L_{\rm [\ion{C}{ii}]}/L_{\rm IR}$ ratio, $\sim 9.4\times10^{-4}$, is consistent with sources discussed in the literature. Our findings show the gas-poor radio AGN hosts have nearby gas-rich companions. We propose that these companions may be stripped clouds resulting from merger processes, which could be a trigger of radio-loud AGN. They may also be a signature of negative AGN feedback (e.g., shock heating) on these infalling companions and on the host galaxy. In general, our analysis shows that powerful AGN at and before Cosmic Noon are impacting and being impacted by cold gaseous clouds in their circumgalactic or protointracluster media.

\end{abstract}



\section{Introduction} \label{sec:intro}
\end{CJK*}
Active galactic nuclei (AGN) feedback regulating galaxy mass build-up is critical for galaxy evolution \citep[e.g.,][]{Harrison_2024}. Molecular gas, the fuel for star formation, can be ejected, disturbed, and/or heated by various feedback processes, leading to galaxy quenching. What triggers AGN activities is not fully understood, but mergers may play a critical role \citep[e.g.,][]{Chiaberge_2015,Wang_companion}. The class of high-redshift radio galaxies (HzRGs, $z\gtrsim1$, $L_{\rm 500\rm MHz}>10^{27}\,\rm W\,Hz^{-1}$), type-2 AGN with the most powerful radio jets, quasar level AGN activities, and no contamination of the central point source due to central obscuration \citep[e.g.,][]{vernet_2001,Miley_2008,Drouart_2014,Falkendal_2019} are ideal sources to examine these processes. Residing in the center of proto-cluster environments and being the progenitors of bright cluster galaxies, HzRGs can help us to understand the evolution of massive systems in dense regions of the Universe \citep[e.g.,][]{Wylezalek_2013,Wylezalek_2014,Noirot_2018}.

Based on spectral energy distribution (SED) fitting \citep[][]{Seymour_2007,DeBreuck_2010,Falkendal_2019}, it has been shown that the massive, $M_{\star}\sim10^{11}\,M_{\odot}$, host galaxies of these radio AGN are quenched or in the process of being quenched. Further observations focusing on ionized gas on large circumgalactic scales showed evidence of feedback \citep[tens to hundreds of kiloparsecs, e.g.,][]{Nesvadba_2006a,Nesvadba_2017a,Nesvadba_2017b,Wang_2021,wang2023}. \citet{kolwa2023} reported that the observed [\ion{C}{i}] (1-0) (a tracer for H$_{2}$, $\nu_{\rm rest}=492.16\,$GHz, hereafter [\ion{C}{i}]) emission is faint in a sample of HzRGs suggesting low content of cold neutral and molecular gas. It remains unclear whether the [\ion{C}{i}] traced H$_{2}$ gas is located within the host galaxy under $\sim2\arcsec$ resolution, if a brighter emission is detected. For example, some well-known HzRGs are found to be gas-rich and undergoing starbursts, but much of the gas and elevated star formation activity is associated with nearby companions revealed in zoom-in views \citep[e.g.,][]{Emonts_2015,ZhongYuxing_2024,Nesvadba_2020}. Therefore, observations at higher resolution and higher sensitivity will help reveal the locations of gas and star formation activity, leading to a better understanding of the role of AGN \citep[e.g.,][]{Wang_2024}.

Combining resolution-matched \textit{JWST}/NIRSpec IFU and Atacama Large Millimeter Array (ALMA) observations, \citet{Wang_companion} reported the ubiquitous detection of gaseous companion systems around $z\approx3.5$ radio AGN. Following up on this discovery, in this Letter, we analyze these companion systems focusing on the ones detected to have cold gas. [\ion{C}{ii}]$158\,\rm \mu m$ ($\nu_{\rm rest}=1900.537\,$GHz, hereafter [\ion{C}{ii}]) is the brightest far infrared (FIR) emission line and an important gas coolant in galaxies \citep[][]{Lagache_2018}. Directly associated with photodissociation region (PDR), the low ionization potential of C atom, 11.3$\,$eV, makes it a tracer of multiple gas phases. Although it is not a tracer of H$_2$ in the presence of hard radiation fields of AGN, this fine structure line is critical for studying regular gas motion, stellar and AGN-driven outflows, and shocks \citep[e.g.,][]{Appleton_2017,Zanella_2018,Ginolfi_2020_outflow,Dessauges-Zavadsky_2020,Lelli_2021,Decarli_2023}. In addition, the FIR continuum under or around [\ion{C}{ii}] offers a direct measure of star formation rate \citep[SFR, e.g.,][]{Falkendal_2019,Schaerer_2020}. Hence, we will study the cold gas content, dynamical masses, and star formation in these companions using the $400\,\rm GHz$ ALMA observations with the assistance of Band 3 and \textit{JWST} data. For this Letter, we assume a flat $\Lambda$ cosmology with $H_{0} = 70\, \rm{km\,s^{-1}\,Mpc^{-1}}$ and $\Omega_{m}=0.3$. Following this luminosity distance, $\rm{1\,arcsec=7.25-7.32\, kpc}$ at the redshifts of this work.

\section{Sample and data} \label{sec:data}
We analyze eight gaseous systems surrounding four $z\approx3.5$ radio galaxies detected in \citet{Wang_companion} with [\ion{C}{ii}]. These HzRGs are selected to have powerful radio jets, rich ancillary data sets, but have different jet morphologies and different host properties \citep[][]{Seymour_2007, DeBreuck_2010,Nesvadba_2007a,Nesvadba_2017a,Nesvadba_2017b,kolwa2023,Wang_2024}.

In this Letter, we primarily use ALMA Band 8 data, focusing on the [\ion{C}{ii}] and continuum (2021.1.00576S, PI: Wuji Wang). The detailed description of the data processing can be found in \citet{Wang_companion}.  We provide a brief summary here. ALMA Band 8 targets the observed frequency at $\sim400\,\rm GHz$ ($\sim158\,\rm \mu m$ in rest frame). We cleaned the ALMA data via the task \texttt{tclean} using the Common Astronomy Software Applications for Radio Astronomy v6.6.0 \citep[CASA,][]{CASA_2022} with Briggs weighting of robust$=+0.5$ for the two targets with 1.7 hours of exposure each (TN J0121+1320 and 4C+19.71) and  robust$=+2.0$ for the other two with 0.8 hours of exposure each (TN J0205+2242 and 4C+03.24). The final [\ion{C}{ii}] cubes, constructed with two spectral windows combined, have $\Delta v=14\,\rm km\,s^{-1}$ per channel. In this Letter, we use the FIR continuum maps from single spectral window for \object{TN J0121+1320} and \object{4C+19.71} (rms noise $76\,\rm \mu Jy\,beam^{-1}$ and $92\,\rm \mu Jy\,beam^{-1}$, respectively). The $3\sigma$ continuum upper limits of \object{TN J0205+2242} and \object{4C+03.24} are obtained from all four spectral window combined, $0.28\,\rm mJy\,beam^{-1}$ and $0.25\,\rm mJy\,beam^{-1}$, respectively. The synthesized beam sizes resulted from the reduction for the four fields range from $\theta_{\rm400\,GHz}\sim0.14\arcsec\times0.12\arcsec$ to $0.23\arcsec \times 0.18\arcsec$ \citep[][]{Wang_companion}.

We also use the ancillary \textit{JWST}/NIRSpec IFU (GO-1970, PI: Wuji Wang) observations to trace the warm ionized gas (e.g., [\ion{O}{iii}]$\lambda$5007). The data processing is described in \citet{Wang_companion}. Using the G235H/F170LP disperser-filter combination, we achieve a spectral resolution of $85-150\,\mathrm{km\,s^{-1}}$. The NIRSpec IFU data has matched spatial resolution as the Band 8 data with point spread function (PSF) of $\sim0.15\arcsec$ \citep[e.g.,][]{Vayner_2023Q3D,Vayner_2023Q3D_2}. The \textit{JWST} data described here may be obtained from the MAST archive at
\dataset[doi:10.17909/jr3z-xc48]{http://dx.doi.org/10.17909/jr3z-xc48}.

In addition, we also use archival ALMA Band 3 observations of cold molecular gas tracer [\ion{C}{i}] \citep[2015.1.00530.S, PI: Carlos De Breuck, and see][for a description of data processing]{kolwa2023}. Unlike the CO transitions, it is less affected by cosmic rays \citep[e.g.,][]{Papadopoulos_2004,Bisbas_2015,Papadopoulos_2018}. We note that the spatial resolution of this Band 3 data, observed $\sim100\,\rm GHz$, is lower than the above two data sets with a median synthesized Band 3 beam size, $\theta_{100\rm GHz}\sim 2\arcsec \times 1.5\arcsec$.

We include the same archival radio maps as in \citet{Wang_companion}. We note that the radio data of 4C+03.24 was newly observed from UT 2024 December 20 to UT 2025 January 3 \citep[24B-147, PI: Carlos De Breuck, see][for a description of reduction]{Wang_companion}.

\begin{figure*}
    \centering
    \includegraphics[width=0.8\textwidth,clip]{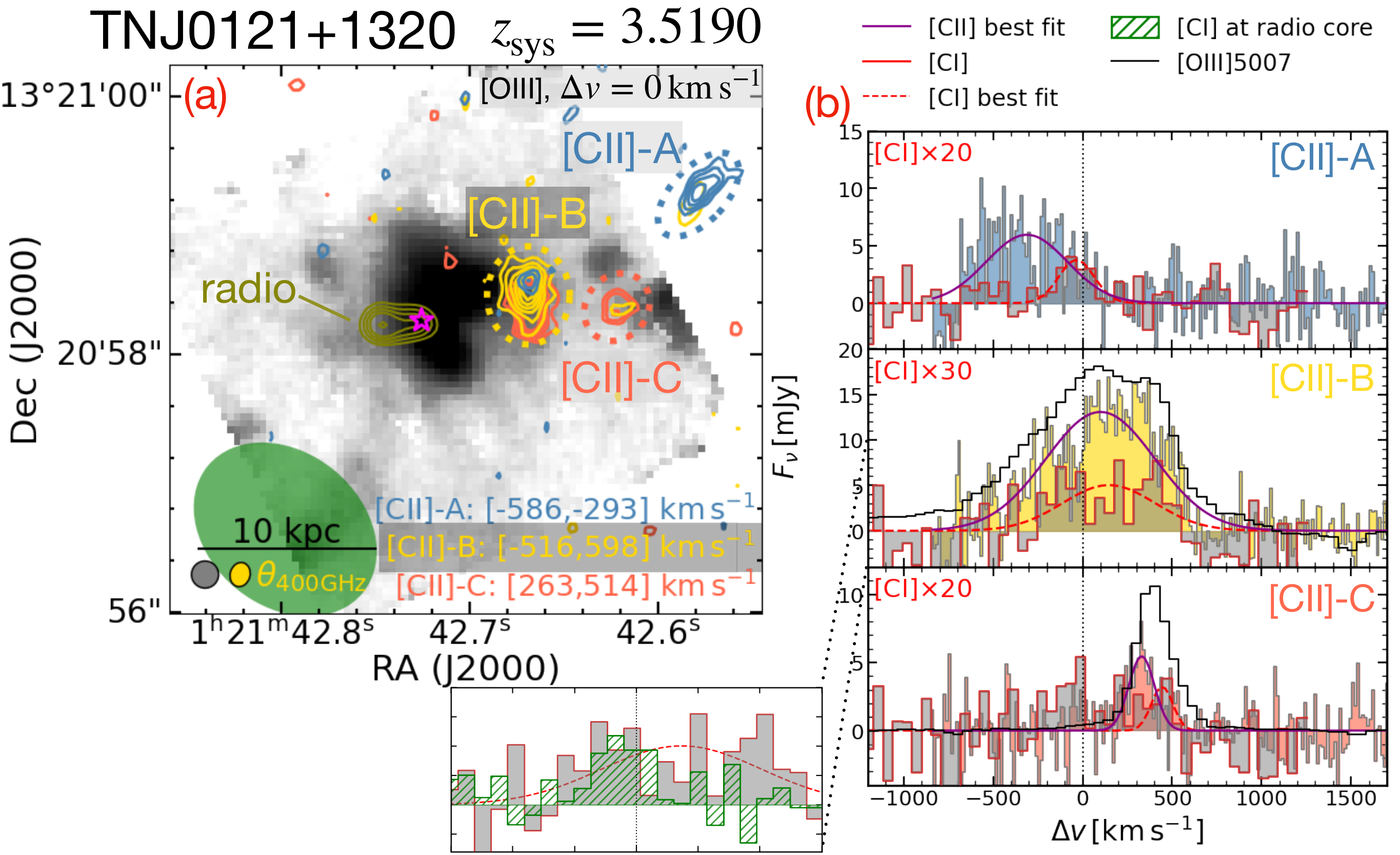}
    \caption{On the left we show [\ion{C}{ii}] moment 0 maps while on the right we present spectra of the companions around TN J0121+1320. \textit{(a)}: Contours of [\ion{C}{ii}] moment 0 maps integrated from velocity channels of each line emission as indicated at the bottom right. The contours are shown in [3, 4, 5, ...]$\sigma$ levels (for [\ion{C}{ii}]-B, only odd levels are shown for better visualization). The gray scale image is [\ion{O}{iii}]$\lambda$5007 surface brightness integrated from three wavelength channels around $\Delta v=0\,\rm km\,s^{-1}$. The color dotted ellipses mark the apertures where the [\ion{C}{ii}] spectra in panel (b) are extracted. The purple star marks the position of the radio core. ALMA Band 8 synthesized beam and representative NIRSpec IFU point spread function are given at the bottom left corner. The VLA X-Band radio map with a synthesized beam size of $0.24\arcsec\times0.24\arcsec$ is shown in dark green contours. \textit{(b)}: Spectra showing [\ion{C}{ii}] emission at the position of the companions. The dark magenta lines show the best single Gaussian fits to the [\ion{C}{ii}] emission. Black histograms are normalized [\ion{O}{iii}]$\lambda$5007 spectra extracted from the same apertures as the [\ion{C}{ii}] spectra. We note that J0121-[\ion{C}{ii}]-A is outside NIRSpec IFU FoV. Red histograms with gray shades are [\ion{C}{i}] spectra at the same positions as the [\ion{C}{ii}] emitters but from aperture with the size of the Band 3 beam size, $\theta_{100\,\rm GHz}=1.98\arcsec\times1.47\arcsec$ (green ellipse). We show the single Gaussian fits to the [\ion{C}{i}] lines in red dashed curve. In the zoom-in panel, [$-600$, 600]$\,\rm km\,s^{-1}$, we compare the [\ion{C}{i}] spectrum extracted at radio core to [\ion{C}{i}] at J0121-[\ion{C}{ii}]-B.}
    \label{fig:tnj0121}
\end{figure*}

\begin{figure*}
    \centering
    \includegraphics[width=0.8\textwidth,clip]{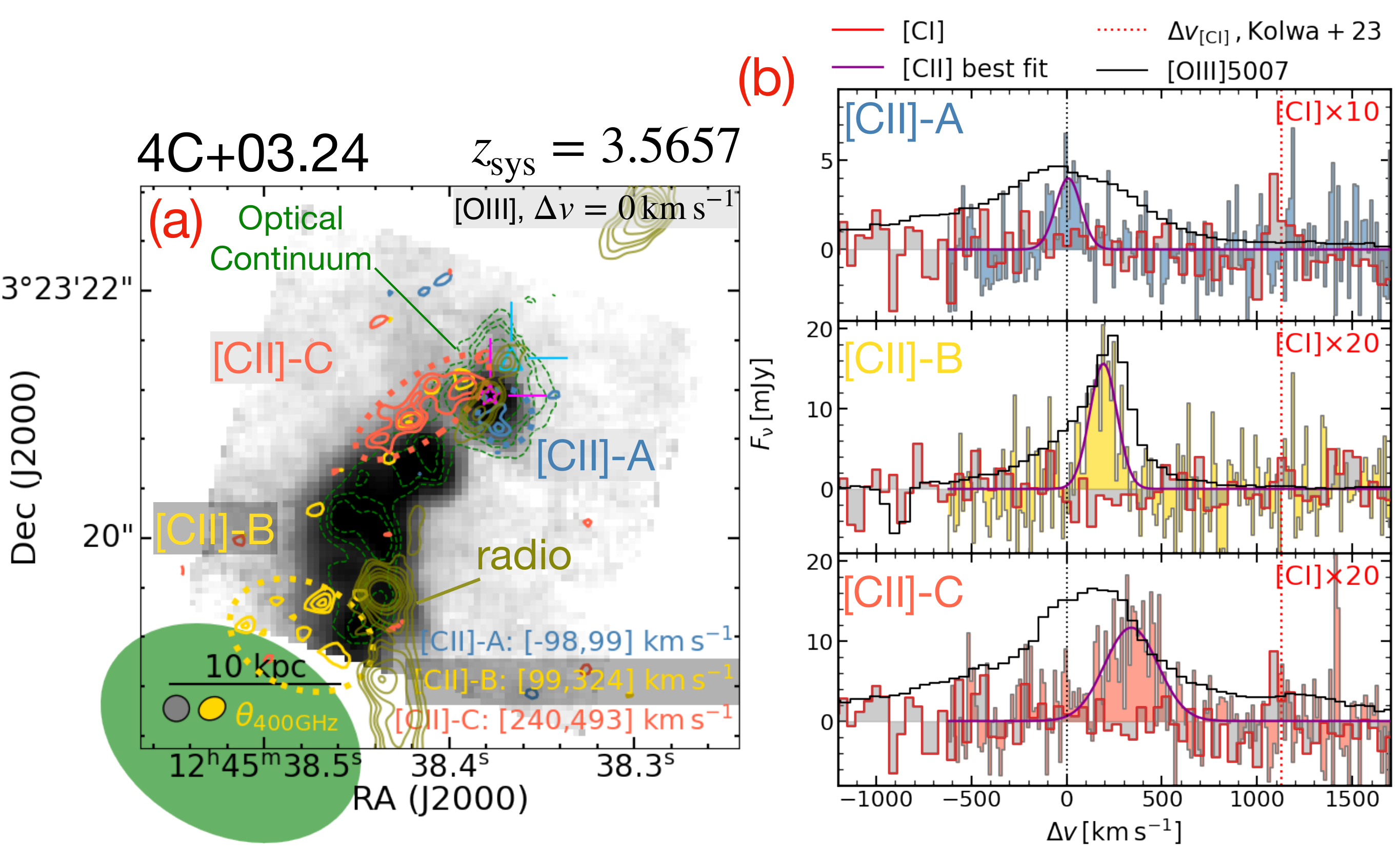}
    \caption{Similar as Fig. \ref{fig:tnj0121}, [\ion{C}{ii}] moment 0 map and spectra of companions of 4C+03.24. \textit{(a)}: The dotted green contours are the optical continuum emissions detected by NIRSpec IFU. The light blue triangle marks the position of [\ion{O}{iii}] companion, 4C03-[\ion{O}{iii}], \citep[][]{Wang_companion} where the spectrum in Fig. \ref{fig:4C03-OIII} is extracted. The VLA Ku-Band radio map with a synthesized beam size of $0.24\arcsec\times0.24\arcsec$ is shown in dark green. \textit{(b)}: The vertical red dotted line indicates the velocity shift of [\ion{C}{i}] detection reported in \citet{kolwa2023}, $\Delta v_{\rm [\ion{C}{i}],K23}=1123\,\rm km\,s^{-1}$, with respect to $z_{\rm sys}=3.5657$. [\ion{C}{i}] spectra are extracted at the same positions as the [\ion{C}{ii}] companions from an aperture with the size of one Band 3 beam, $\theta_{100\,\rm GHz}=2.23\arcsec\times1.62\arcsec$ (green ellipse).}
    \label{fig:4c03}
\end{figure*}

\begin{figure*}
    \centering
    \includegraphics[width=0.8\textwidth,clip]{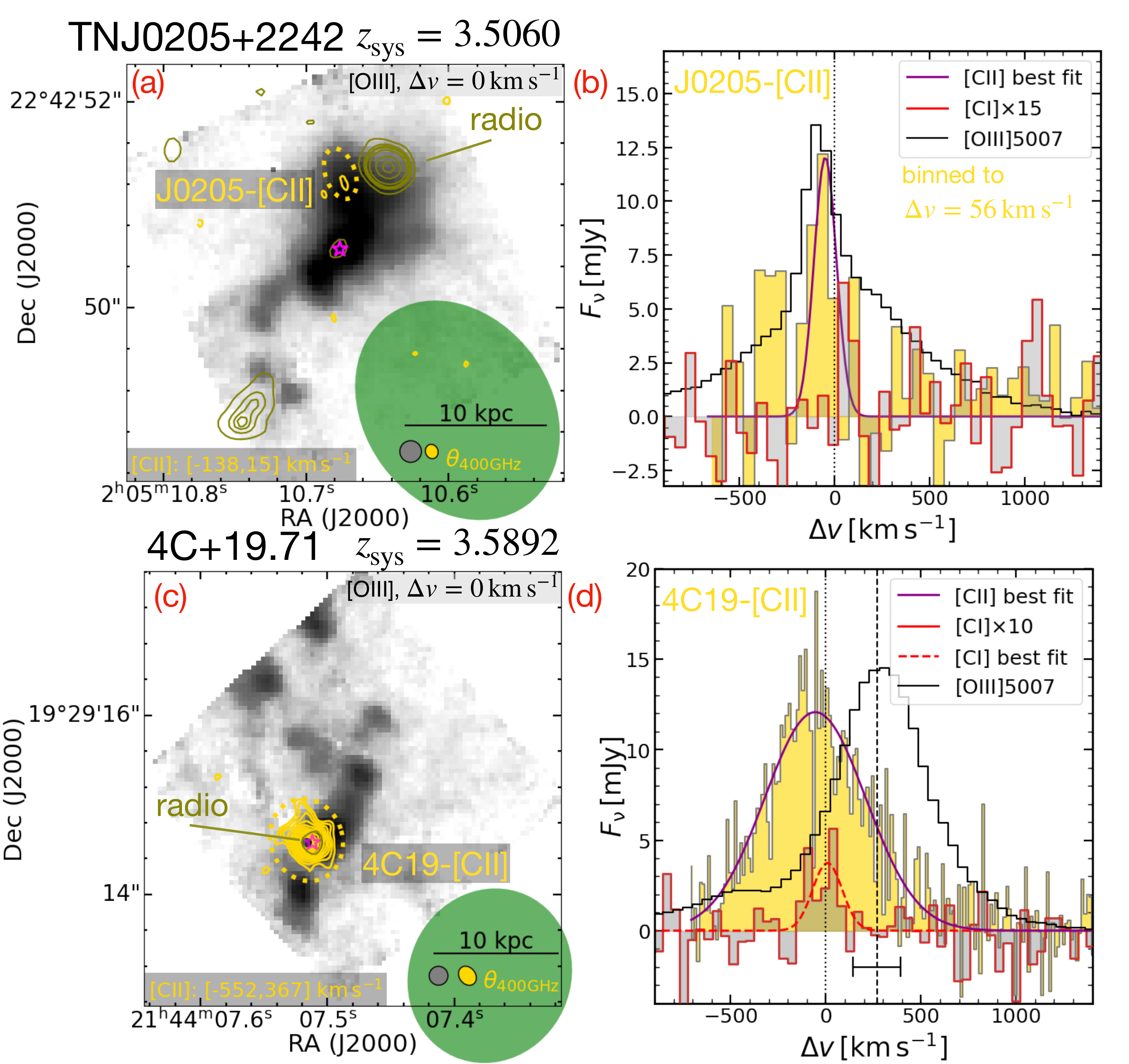}
    \caption{Similar as Fig. \ref{fig:tnj0121}, [\ion{C}{ii}] moment 0 maps and spectra of the companions of TN J0205+2242 (\textit{a,b}) and 4C+19.71 (\textit{c,d}). VLA X-Band ($0.20\arcsec\times0.20\arcsec$) and C-Band ($0.25\arcsec\times0.24\arcsec$) radio maps are shown for TN J0205+2242 and 4C+19.71, respectively. The vertical dashed black line in panel (d) marks the $\Delta v_{[\ion{O}{iii}]}=268\,\rm km\,s^{-1}$ ($z_{[\ion{O}{iii}]}=3.5933$ which might be the systemic redshift of the radio AGN, see Sect. \ref{sec:analysis}). The horizontal black bar indicates the $250\,\rm km\,s^{-1}$ range where the $L_{[\ion{C}{ii}]}$ upper limit at radio core is integrated. For TN J0205+2242, the [\ion{C}{i}] spectrum is extracted from an aperture with the size of one Band 3 beam, $\theta_{100\,\rm GHz}=2.28\arcsec\times1.81\arcsec$ (green ellipse), and is further binned by a factor of four, i.e., $56\,\rm km\,s^{-1}$ per channel. For 4C+19.71, its [\ion{C}{i}] spectrum is extracted from an aperture with the size of one Band 3 beam, $\theta_{100\,\rm GHz}=1.98\arcsec\times1.81\arcsec$ (green ellipse). Only the radio core of 4C+19.71 is shown where the lobes are outside the FoV \citep[][]{Wang_2024}.}
    \label{fig:j0205_4c19}
\end{figure*}

\begin{table*}
 \caption{Emission line fitting results and inferred properties of our [\ion{C}{ii}] compaions.}\label{tab:results}
 \centering
\begin{tabular}{c c c c c c c c c c}
\hline
\hline
\colhead{Companion}  & $\Delta v_{\rm [\ion{C}{ii}]}$ & $\sigma_{v,\rm[\ion{C}{ii}]}$ & $I_{\rm [\ion{C}{ii}]}$  & \multicolumn{3}{c}{Luminosity $L/L_{\odot}$} & $M_{\rm H_{2},[\ion{C}{i}]}/M_{\odot}$ & SFR$_{\rm IR}$ & $M_{\rm dyn}^{\rm dis}/M_{\odot}$  \\
    &  [$\rm km\,s^{-1}$]  & [$\rm km\,s^{-1}$]  &  [$\rm Jy\, km\,s^{-1}$]  & $\times10^{9}$ & $\times10^{12}$ & $\times10^{6}$   & $\times10^{10}$  & [$M_{\odot}\rm\,yr^{-1}$] & $\times10^{10}$ \\
\multicolumn{1}{c}{(1)} & (2) & (3) & (4) & \multicolumn{3}{c}{(5)} & (6) & (7) & (8) \\
\cmidrule(lr){5-7} 
 &  &   &  &  [\ion{C}{ii}] & IR & [\ion{C}{i}] & & &\\

\hline
J0121-[\ion{C}{ii}] & & & & & & & & & \\
$z_{\rm sys}=3.5190$ & & & & $<0.20^{\dag}$& & $5.5\pm2.5^{\ddag}$ & $1.0\pm0.6^{\ddag}$ & & \\
\cline{1-1}
A   & -312$\pm$28 & 226$\pm$29 & 3.38$\pm$0.37 & 1.40$\pm$0.15 & 1.54$\pm$0.21 & $4.3\pm2.0^{\ddag}$ & $1.0\pm0.5^{\ddag}$  & 156$\pm$63 & 12 \\

B   & 100$\pm$38  & 305$\pm17$ & 9.98$\pm$0.47 & 4.13$\pm$0.20 & 6.89$\pm$0.22 & 11.6$\pm$3.4          & 2.7$\pm0.8$         & 696$\pm$66 & 26(9.0)$^{\S}$ \\

C   & 330$\pm$13  & 72$\pm$13  & 0.98$\pm$0.16 & 0.41$\pm$0.06 & $<$0.3        & $<1.9^{\ddag}$ & $<0.4^{\ddag}$  & $<$34      & 1.0 \\
\hline
J0205-[\ion{C}{ii}] & & & & & & & & & \\
$z_{\rm sys}=3.5060$ & & & & $<1.22^{\dag}$& & & & & \\
\cline{1-1}

    & -49$\pm$22 & 59$\pm$22   & 1.78$\pm$0.57  & 0.73$\pm$0.23 & $<$0.7      & $<$5              & $<$1.1              & $<$73      & 1.9 \\

\hline
4C03-[\ion{C}{ii}] & & & & & & & & & \\
$z_{\rm sys}=3.5657$ & & & & $<0.44^{\dag}$& & & & & \\
\cline{1-1}
A   & 9$\pm$16   & 66$\pm$17  & 0.67$\pm$0.14  & 0.28$\pm$0.06  & $<$0.4   & $<$5$^{\ddag}$       & $<$1.3$^{\ddag}$  & $<$35  & 1.9 \\

B   & 193$\pm$11 & 71$\pm$11  & 2.81$\pm$0.38  & 1.19$\pm$0.16  & $<$1.2   & $<$5      & $<$1.2 & $<$125 & 4.5 \\

C   & 340$\pm$19 & 139$\pm$19 & 4.08$\pm$0.49  & 1.73$\pm$0.21  & $<$0.9   & $-^{\ddag}$      & $-^{\ddag}$  & $<$92  & 13  \\
\hline
4C19-[\ion{C}{ii}] & & & &  & & & & &  \\
$z_{\rm sys}=3.5892$ & & & & $<0.55^{\dag}$ & & & & & \\
\cline{1-1}

    & -56$\pm$11 & 266$\pm$11 & 8.04$\pm$0.30 & 3.44$\pm$0.13  & 4.45$\pm$0.18  & 8$\pm$3 & 1.9$\pm$0.7 & 449$\pm$18  & 29(7.0)$^{\S}$ \\
 \hline
\end{tabular}
\parbox{\textwidth}{
\smallskip
\textbf{Notes:} 
(1) Name of the companion and systemic redshift, $z_{\rm sys}$, following \citet{Wang_companion}. (2) -- (4) Fitted [\ion{C}{ii}] velocity shift ($\Delta v_{\rm [\ion{C}{ii}]}$ with respect to $z_{\rm sys}$), line width ($\sigma_{v,\rm[\ion{C}{ii}]}$ uncorrected for ALMA spectra resolution), and intensity ($I_{\rm [\ion{C}{ii}]}$). (5) Emission lines, [\ion{C}{ii}] and [\ion{C}{i}], and total IR luminosities in $L_{\odot}$. (6) Molecular gas mass converted from [\ion{C}{i}], $M_{\rm H_{2},[\ion{C}{i}]}$, in $M_{\odot}$. (7) Star formation rate from $L_{\rm IR}$. (8) Dynamical mass calculated from $\sigma_{v,\rm[\ion{C}{ii}]}$ using Eq. \ref{eq:M_dyn}. The quoted errors in this table are propagated $1\sigma$ measurement errors. {\rm{\dag}~}{Upper limits derived at the radio core position.} {\rm{\ddag}~}{Due to the overlap between [\ion{C}{i}] spectra extraction regions, we correct the reported $L_{\rm [\ion{C}{i}]}$ and $M_{\rm H_{2}},[\ion{C}{i}]$ for J0121-radio core, [\ion{C}{ii}]-A and [\ion{C}{ii}]-C. 4C03-[\ion{C}{ii}]-A overlaps with [\ion{C}{ii}]-C $>60\%$ such that only one upper limit is reported here (see text).} {\rm{\S}~}{For J0121-[\ion{C}{ii}]-B and 4C19-[\ion{C}{ii}], the $M_{\rm dyn}$ assuming rotational disk were given in parentheses \citep[][]{Wang_companion}. } 
}
\end{table*}


\section{Analysis} \label{sec:analysis}
In \citet{Wang_companion}, we reported the ubiquitous discovery of nearby, $\lesssim18\,$kpc, companion systems in the fields around the four $z\approx3.5$ radio AGN. This sample of HzRGs was selected to have diverse properties in terms of, e.g., SFR and radio morphologies, with no requirement that they reside in known over-dense environments. ``Companions'' under their definition represent systems either with (i) peculiar warm ionized gas kinematics different from the bulk gas motion or (ii) a detection of [\ion{C}{ii}] emission. Hence, these companions may be gaseous clouds off-center from the AGN. We primarily focus on the [\ion{C}{ii}] companions in this Letter to analyze the cold gas content. We briefly summarize the selection here \citep[see][for details]{Wang_companion}. The [\ion{C}{ii}] companions are selected first based on visual check in the field of ALMA Band 8 primary beam ($\sim10\arcsec\times10\arcsec$) over the spectral bandwidth of $\sim5300\,\rm km\,s^{-1}$. The companions are then checked to have a signal-to-noise ratio (S/N) $\gtrsim3$ collapsed over $125\,\rm km\,s^{-1}$ and are validated spectrally. Following the naming convention in \citet{Wang_companion}, the companions are named by the central radio AGN following the detection method.

We show the [\ion{C}{ii}] moment 0 intensity contours in Fig. \ref{fig:tnj0121}a, \ref{fig:4c03}a, and \ref{fig:j0205_4c19}ac overlaid on [\ion{O}{iii}] surface brightness maps detected by NIRSpec IFU. We find that the [\ion{C}{ii}] and [\ion{O}{iii}] exhibit different spatial distributions, indicating that the cold and warm gas originate from distinct clouds. For example, [\ion{O}{iii}] closely traces AGN ionization, whereas [\ion{C}{ii}] traces companions outside AGN hosts. The [\ion{C}{ii}] spectra are shown in Fig. \ref{fig:tnj0121}b, \ref{fig:4c03}b, and \ref{fig:j0205_4c19}bd. To maximize the S/N of the [\ion{C}{ii}] detections, we customize the extraction apertures based on their morphology in the moment 0 maps (color dotted ellipses). We then perform single-Gaussian fitting to [\ion{C}{ii}] line. 

It is clear that the gas traced through [\ion{C}{ii}] emission is not located at the position of the radio galaxy based on their distributions. To quantitatively compare the cold gas content between companions and host galaxies, we estimate the [\ion{C}{ii}] upper limits at the radio core by assuming a disk size of 5~kpc \citep[i.e., $\sim0.7\arcsec$ as the extraction aperture diameter, e.g.,][]{Venemans_2020}. For TN J0121+1320 and TN J0205+2242 where the radio core is $>5\,$kpc away from the [\ion{C}{ii}] companions, we take the $3\sigma$ [\ion{C}{ii}] intensity upper limit from a velocity range of [$-125$, 125]$\,\rm km\,s^{-1}$. This range equals to the median full width at half maximum (FWHM) of all fitted [\ion{C}{ii}] lines (Table \ref{tab:results}). We note that the [\ion{C}{ii}] line width at radio core may be expected to be broader due to feedback and/or due to beam smearing of the inner rotation curve of a massive galaxy. If we double the velocity range, i.e., two times of FWHM, the estimated $L_{\rm [\ion{C}{ii}]}$ upper limit at radio core would be $\sim30\%$ higher. For 4C+03.24 where companion [\ion{C}{ii}] emissions are near the radio core, we simply integrate the line intensity from [$-125$, 125]$\,\rm km\,s^{-1}$ as the upper limit. As discussed in \citet{Wang_companion}, we conclude that 4C19-[\ion{C}{ii}] is a separate system from the AGN host: (i) its [\ion{O}{iii}] emission peak, i.e., ionized gas tracing systemic redshift of AGN, has a shift of $\Delta v_{[\ion{O}{iii}]}=268\,\rm{km\,s^{-1}}$ from the [\ion{C}{i}] based redshift ($\Delta v=0\,\rm{km\,s^{-1}}$ in Fig. \ref{fig:j0205_4c19}d); (ii) the spatial separation only suggests a $<5\%$ probability that the [\ion{C}{ii}] coincides with the core \citep[see a discussion in][]{Wang_companion}. Here, we integrate the [\ion{C}{ii}] spectrum at the radio core in [$-125+\Delta v_{[\ion{O}{iii}]}$, $125+\Delta v_{[\ion{O}{iii}]}$]$\,\rm km\,s^{-1}$ as the upper limit in the AGN host of 4C+19.71 (horizontal bar in Fig. \ref{fig:j0205_4c19}d).

 The previous [\ion{C}{i}] analysis by \citet{kolwa2023} focused only on radio core positions. With the guide of our new [\ion{C}{ii}] detections at higher spatial resolution, we re-extract the [\ion{C}{i}] spectra at the same position as [\ion{C}{ii}] companions using apertures corresponding to the Band 3 synthesized beam size (Fig. \ref{fig:tnj0121}b, \ref{fig:4c03}b, and \ref{fig:j0205_4c19}bd). Given the new data, other sources of contamination in [\ion{C}{i}] detection are expected to be negligible, e.g., tentative emissions at offset redshifts and positions. We then perform single-Gaussian fitting to the [\ion{C}{i}] line when detected. In case of a non-detection, we calculate the $3\sigma$ [\ion{C}{i}] intensity upper limit from a velocity range of [$-125$, 125]$\,\rm km\,s^{-1}$ following the estimate of [\ion{C}{ii}] upper limits. We caution that for fields of TN J0121+1320 and 4C+03.24 where multiple [\ion{C}{ii}] companions are detected, the extraction apertures of [\ion{C}{i}] may be overlapped. Specifically, J0121-[\ion{C}{ii}]-C has a $\sim28\%$ overlap in area with [\ion{C}{ii}]-B and a $\sim15\%$ overlap with [\ion{C}{ii}]-A, respectively. We also extract [\ion{C}{i}] at radio core which has a $\sim15\%$ overlap with [\ion{C}{ii}]-B. Since the brightest [\ion{C}{i}] emission is detected at [\ion{C}{ii}]-B, we correct [\ion{C}{i}] at [\ion{C}{ii}]-C and radio core for overlap, $f_{\rm [\ion{C}{i}]} = (1-P_{\rm overlap})f_{\rm [\ion{C}{i}],fit}$, where $P_{\rm overlap}$ is the percentage of overlap. We take [\ion{C}{i}] at [\ion{C}{ii}]-C as an upper limit given its S/N$\lesssim3$. Since [\ion{C}{ii}]-A does not overlap with [\ion{C}{ii}]-B, we take it as a detection after correcting for overlap (see discussion in Sect. \ref{subsec:dis_cold}). 4C03-[\ion{C}{ii}]-A has an overlap with [\ion{C}{ii}]-C $>60\%$ such that we only report a single upper limit estimate in [\ion{C}{i}] for these two.

We compute the emission line luminosity, $L_{\rm line}$, with the equation \citep[][]{Carilli_2013,Bethermin_2020,Falkendal_2021}:
\begin{equation}\label{eq:L_line}
    \frac{L_{\rm line}}{L_{\odot}} = 1.04\times10^{-3}\times\left[ \frac{I_{\rm line}}{\rm Jy\,km\,s^{-1}}  \right] \left[\frac{D_{\rm L}}{\rm Mpc}\right]^{2} \left[\frac{\nu_{\rm obs, line}}{\rm GHz}\right]
    ,
\end{equation}
where $I_{\rm line}$ is the line intensity, $D_{\rm L}$ is the luminosity distance in Mpc, and $\nu_{\rm obs, line}$ is the observed line frequency. 

In the radiation field of bright quasars, the relation between [\ion{C}{ii}] and $M_{\rm H_{2}}$ calibrated for star forming galaxies may not hold \citep[e.g.,][]{Zanella_2018,Dessauges-Zavadsky_2020,Decarli_2023}. The neutral carbon fine structure transitions are found to be a good tracer of cold molecular gas in these extreme fields \citep[or at least tracing cold neutral gas, e.g.,][]{Papadopoulos_2004,Huang_2024,Emonts_2023}. Without additional data on higher order [\ion{C}{i}] transitions,  we calculate $M_{\rm H_{2}}$ solely based on [\ion{C}{i}](1-0) following \citet{kolwa2023}: 
\begin{multline}\label{eq:m_h2}
    \frac{M_{\rm H_{2},[\ion{C}{i}]}}{M_{\odot}} = \frac{1375.8}{Q_{10}(1+z)} \times  \\
    \left[\frac{D_{\rm L}}{\rm Mpc}\right]^{2}\left[\frac{X_{\rm[\ion{C}{i}]}}{10^{-5}}\right]^{-1} \left[ \frac{A_{10}}{10^{-7}\,{\rm s^{-1}}}\right]^{-1} \left[ \frac{I_{\rm [\ion{C}{i}]}}{\rm Jy\,km\,s^{-1}}  \right],
\end{multline}
where the excitation factor $Q_{10}$ is assumed as $Q_{10}=0.48$, the [\ion{C}{i}]-to-$\rm H_{2}$ ratio is assumed as $X_{[\ion{C}{i}]}=3\times10^{-5}$, and $A_{10}$ is the Einstein A-coefficient with $A_{10}=7.93\times10^{-8}\,\rm s^{-1}$ \citep[][]{Weiss_2003, Papadopoulos_2004,Falkendal_2021}.

If the [\ion{C}{ii}] emitting gas clouds are virialized, we can estimate the dynamical masses of these systems assuming that they are dispersion-dominated \citep[i.e., $v_{\rm rot}/\sigma_{v}\lesssim\sqrt{3.36}$, where $v_{\rm rot}$ is the rotation velocity,][]{ubler_2023, Dessauges-Zavadsky_2020,Forster-Schreiber_2020}:  
\begin{equation}\label{eq:M_dyn}
    M_{\rm dyn}^{\rm dis} = K(n)K(q)\frac{\sigma_{v,\rm[\ion{C}{ii}]}^{2}r_{e}}{G}
\end{equation}
where $G$ is the gravitational constant, $r_{e}$ is the effective radius, $\sigma_{v,\rm[\ion{C}{ii}]}$ is the fitted Gaussian line width (after spectral resolution correction), $K(n)=8.87-0.831n+0.0241n^{2}$ is a function of S\'{e}rsic index, $n$, and $K(q)=[0.87+0.38e^{-3.71(1-q)}]^{2}$ is a function of axis ratio, $q$. We perform a 2D Gaussian fit to the [\ion{C}{ii}] moment 0 maps using \texttt{lmfit} task in CASA and take $r_{e}=\sqrt{\rm FWHM_{x}\times FWHM_{y}}/2$, where FWHM$_{x}$ and FWHM$_{y}$ are fitted Gaussian widths after deconvolution with the beam. For J0205-[\ion{C}{ii}], we use $r_{e}=2\sqrt{\theta_{\rm 400\,GHz}}$. Since $K(q)$ is monotonically increasing with $q\in[0,1]$ and $K(n)$ is monotonically decreasing with $n\in[1,8]$, we take $q=1$ and $n=1$ in this calculation which returns the upper limit of $M_{\rm dyn}^{\rm dis}$ \citep[][]{ubler_2023,Maiolino_2023}. In \citet{Wang_companion}, we estimated the $M_{\rm dyn}$ assuming disk rotation using $^{\rm 3D}$\textsc{Barolo} for J0121-[\ion{C}{ii}]-B and 4C19-[\ion{C}{ii}] \citep[Table \ref{tab:results},][]{DiTeodoro_2015}. 

 FIR continuum is known to be related to star formation \citep[][]{MandD_2014,Falkendal_2019, Bethermin_2020,Fudamoto_2020}. We extract the continuum flux at the same position and from the same aperture as the [\ion{C}{ii}] emissions. Following \citet{Bethermin_2023,Fudamoto_2020}, we convert the monochromatic luminosity into the total IR luminosity (integrated in $8-1000\,\rm \mu m$) using $L_{\rm 158\,\mu m}=L_{\rm IR}/0.13$ and then to SFR with ${\rm SFR}_{\rm IR}=2.64\times10^{-44}L_{\rm IR} [\frac{M_{\odot}\,{\rm yr}^{-1}}{\rm erg\,s^{-1}}]$ \citep[][]{MandD_2014}. We note that we are observing at the peak frequency of cold dust emission where the contamination of hotter dust, $T\gtrsim60\,\rm K$, in HzRGs is known to be negligible \citep[see][]{DeBreuck_2010,Falkendal_2019}. For the non-detections, we calculate $3\sigma$ Band 8 continuum upper limits. For J0205-[\ion{C}{ii}] and 4C03-[\ion{C}{ii}], the $3\sigma$ $L_{\rm IR}$ upper limits are derived from Band 8 continuum maps using all four spectral windows assuming equal contribution. For J0121-[\ion{C}{ii}]-C, the $L_{\rm IR}$ upper limit is from a single spectral window \citep[][]{Wang_companion}. In the following discussion, we use the SFR$_{\rm IR}$ derived above. \citet{Falkendal_2019} reported SFR$_{\rm SED}$ for our sample based on SED fitting where the main photometric constraints come from low spatial resolution \textit{Herschel}/SPIRE and JCMT/SCUBA \citep[$\sim20\arcsec$,][]{Archibald_2001,Drouart_2014}. Our SFR$_{\rm IR}$ estimate in this work is consistent with \citet{Falkendal_2019} SED fitting measurements of TN J0121+1320, ${\rm SFR_{SED}}=626_{-243}^{+267}\,M_{\odot}\,{\rm yr}^{-1}$. Our SFR$_{\rm IR}$ upper limit estimates for the two targets with shallow exposure  are also consistent with their ${\rm SFR_{SED}}$ ($<84\,\,M_{\odot}\,{\rm yr}^{-1}$ and $142_{-130}^{+238}\,\,M_{\odot}\,{\rm yr}^{-1}$ for TN J0205+2242 and 4C+03.24, respectively).  For 4C+19.71, our SFR$_{\rm IR}$ is higher by a factor of $\sim2$ compared to its $\mathrm{SFR_{\rm SED}}=84_{-62}^{+176}\,M_{\odot}\,{\rm yr}^{-1}$ where only JCMT/SCUBA shows photometry detections \citep{Falkendal_2019}. The discrepancy might be raised from the difference in spatial resolution. 
 
\begin{figure*}
    \centering
    \includegraphics[width=0.8\textwidth,clip]{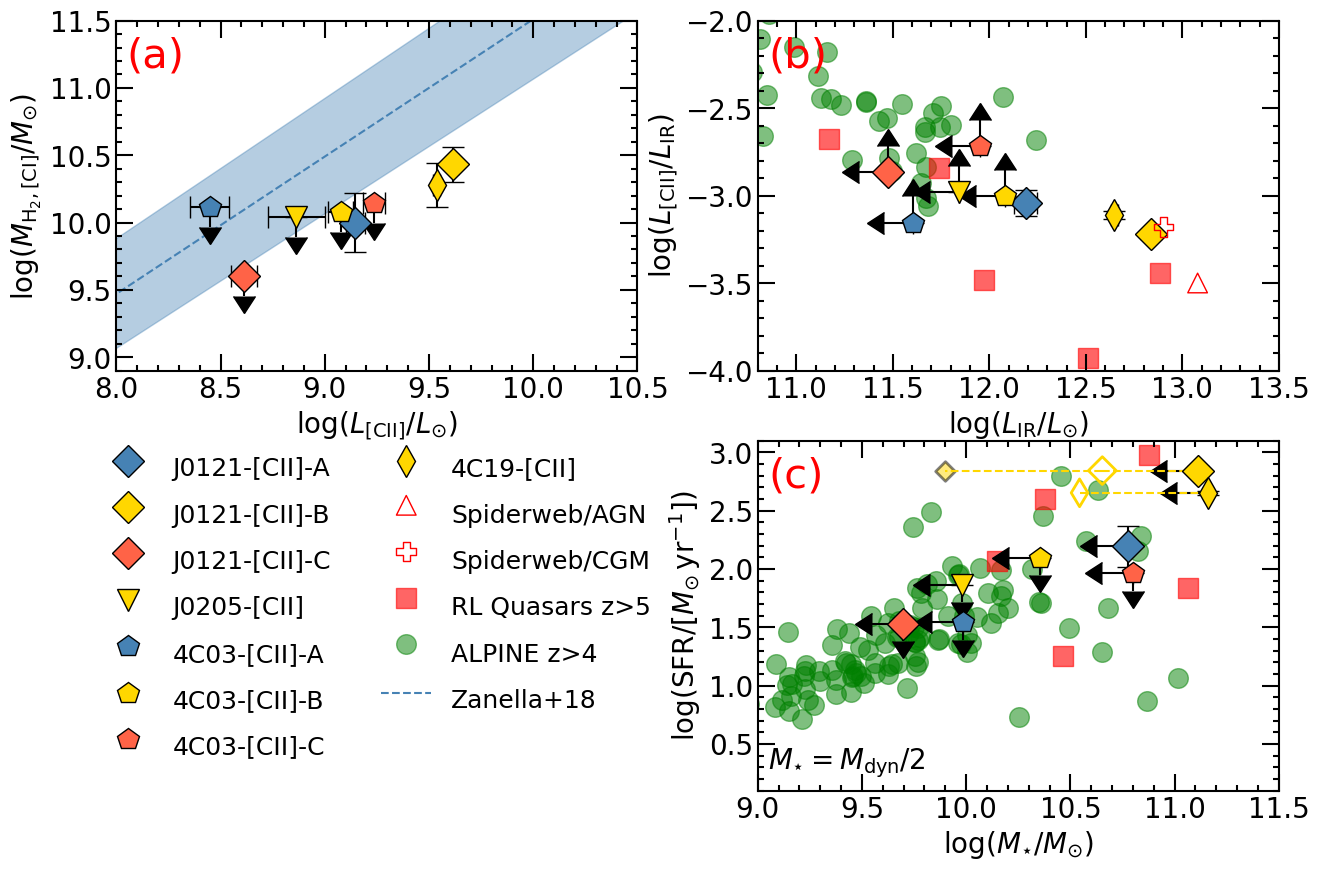}
    \caption{Sample properties and comparison: \textit{(a)} $L_{\rm [\ion{C}{ii}]}$ versus $M_{\rm H_{2}, [\ion{C}{i}]}$, \textit{(b)} $L_{\rm IR}$ versus $L_{\rm [\ion{C}{ii}]}/L_{\rm IR}$, and \textit{(c)} $M_{\star}$ versus SFR. In panel (a), we show the $M_{\,\rm H_{2}}-L_{\rm [\ion{C}{ii}]}$ relation with uncertainties \citep[][]{Zanella_2018}. For comparison in panel (b) and (c), we include $z>5$ radio-loud quasar hosts measurements \citep[red filled squares, ][]{Mazzucchelli_2025}. We also include measurements of $4<z<6$ star forming main sequence galaxies from ALPINE survey \citep[green circles][]{LeFevre_2020,Faisst_2020,Bethermin_2020,Schaerer_2020}. We include $L_{\rm IR}$ measured from CRISTAL survey \citep[i.e., deeper observations covering part of ALPINE targets, see e.g.,][]{Mitsuhashi_2024}. We also add measurements of Spiderweb (open red markers, $z=2.16$ radio galaxy) \citep[panel (b)][]{Seymour_2012, DeBreuck_2022}. 
    The SFR of ALPINE galaxies used here are from SED fitting. Our $M_{\star}$ are estimated assuming $M_{\star}=M_{\rm dyn}/2$ (see text). We also give the $M_{\rm dyn}$ of 4C19-[\ion{C}{ii}] and  J0121-[\ion{C}{ii}]-B (also a rough direct $M_{\star}\sim10^{9.9}\,M_{\odot}$ estimation from optical continuum in small marker) derived based on rotational disk fit in open symbols \citep[][]{Wang_companion}.
    }
    \label{fig:relation}
\end{figure*}

We summarize the derived properties of our eight [\ion{C}{ii}] companions in Table \ref{tab:results}. In Fig. \ref{fig:relation}, we present the relations between these properties, along with comparisons to other high-$z$ samples. Since our data only probes the inner $\sim3-8\,$kpc of these companions, we do not assume that the dark matter takes a large fraction in $M_{\rm dyn}$. Adopting a gas mass fraction of $\sim1$ \citep[$M_{\rm gas}/(M_{\rm gas}+M_{\star})$, e.g.,][]{Geach_2011,Lehnert_2016}, we estimate zeroth-order stellar mass estimates, $M_{\star}=M_{\rm dyn}/2$, and show them in Fig. \ref{fig:relation}c. 

\section{Discussion and conclusion} \label{sec:disc}

\subsection{Cold gas clouds around but not at radio AGN}\label{subsec:dis_cold}

The immediate conclusion from our [\ion{C}{ii}] detections is that there are cold gas systems in the surroundings of $z\approx$3.5 radio AGN, but the majority of the gas is not at the AGN position. The AGN positions are determined from the cores of the radio maps. The systemic redshifts are derived based on \ion{He}{ii}$\lambda1640$ for TN J0121+1320 and TN J0205+2242 or [\ion{C}{i}] if \ion{He}{ii}$\lambda1640$ is unavailable for 4C+19.71 \citep[$z_{\rm sys}$ from VLT/SINFONI \textnormal{[\ion{O}{III}]}$\lambda5007$ was used for 4C+03.24, see][for more discussions]{kolwa2023,Wang_companion}. Our companions have $2.8\times10^{8}<L_{[\ion{C}{ii}]}/L_{\odot}<4.2\times10^{9}$ with a median of $1.3\times10^{9}\,L_{\odot}$. $L_{[\ion{C}{ii}]}$ upper limits at the radio core are estimated to be $\lesssim10^{9}\,L_{\odot}$ (Sect. \ref{sec:analysis} and Table \ref{tab:results}).  As shown in Fig. \ref{fig:tnj0121}b, \ref{fig:4c03}b, and \ref{fig:j0205_4c19}b, the [\ion{C}{ii}] companions can have, in three cases, similar kinematic counterparts as in [\ion{O}{iii}]$\lambda$5007 (J0121-[\ion{C}{ii}]-C, 4C03-[\ion{C}{ii}]-B, and J0205-[\ion{C}{ii}]). This indicates that the companions are in the same potential well as the radio AGN host. The [\ion{C}{ii}] line widths are generally narrower than the pure ionized gas tracer with a median $\sigma_{v}=105\,\rm km\,s^{-1}$ and most of the [\ion{C}{ii}] emitting clouds are spatially unresolved, indicating that the companion clouds are less perturbed by the AGN or they have much lower mass (4C+03.24 may be an exception). 

In TN J0121+1320 and 4C+19.71, we detect [\ion{C}{i}] emissions using extraction apertures centered at the [\ion{C}{ii}] companions \citep[see also][]{kolwa2023}. We find that the two carbon line profiles are in agreement. In Fig. \ref{fig:tnj0121}b zoom-in panel, we compare the [\ion{C}{i}] lines extracted at J0121-[\ion{C}{ii}]-B (red) and radio core (green). Despite the overlap in extraction due to large $\theta_{\rm100\,GHz}$ (Sect. \ref{sec:analysis}),  $L_{[\ion{C}{i}]}$ at J0121-[\ion{C}{ii}]-B is $\sim 2$ times brighter than at the radio core  (Table \ref{tab:results}). This implies that the H$_{2}$ gas (or at least cold neutral gas) is located more likely at the companion position. Taking also the fact that J0121-[\ion{C}{ii}]-C extraction region has a larger overlap with [\ion{C}{ii}]-B ($\sim28\%$ versus $\sim15\%$, Sect. \ref{sec:analysis}) but having fainter $L_{[\ion{C}{i}]}$, this may further imply that the H$_2$ gas is located slightly towards east than west if not solely at [\ion{C}{ii}]-B. A further inference may be that there is a gas exchange between the companion and host. The [\ion{C}{i}] spectrum at J0121-[\ion{C}{ii}]-A is consistent with the broad profile of [\ion{C}{ii}] and is not contaminated by [\ion{C}{ii}]-B (Sect. \ref{sec:analysis}, Fig. \ref{fig:tnj0121}b). It is new evidence of off-center H$_2$ which was missed in \citet{kolwa2023}. As discussed in Sect. \ref{sec:analysis}, cold gas traced by [\ion{C}{ii}] and [\ion{C}{i}] is not in the AGN host of 4C+19.71. The [\ion{C}{i}] detection in 4C+03.24 by \citet{kolwa2023} is reported to be at $\Delta v_{\rm [\ion{C}{i}],K23}=1123\,\rm km\,s^{-1}$ from the ionized gas (Fig. \ref{fig:4c03}). This may be a false detection as no [\ion{C}{ii}] has been detected at the same $\Delta v$ (see  Appendix \ref{app:H2_4C03}). We stress that the main conclusion that the companions are rich in cold gas is based on [\ion{C}{ii}] detections, such that overlap in some of the [\ion{C}{i}] extractions does not affect it. Deeper and higher spatial resolution [\ion{C}{i}] are required to more accurately determine the $M_{\rm H_{2}, [\ion{C}{i}]}$

Literature studies have demonstrated that [\ion{C}{ii}], which is closely associated with PDRs, may serve as a tracer of H$_2$ \citep[e.g.,][]{Zanella_2018, Dessauges-Zavadsky_2020}. We compare the molecular gas mass derived from [\ion{C}{i}], $M_{\rm H_{2}, [\ion{C}{i}]}$, and $L_{[\ion{C}{ii}]}$ for our sample in Fig. \ref{fig:relation}a together with the \citet{Zanella_2018} $M_{\rm H_{2}}-L_{\rm \ion{C}{ii}}$ relation. Three of our detections generally follow the relation while the majority have a tendency to fall below it especially for brighter [\ion{C}{ii}] emitters ($>10^{9}\,L_{\odot}$). This is expected because in the vicinity, $\lesssim18\,$kpc, of bright quasars with the energetic jets, a larger fraction of [\ion{C}{ii}] may be associated with ionized and/or neutral gas than molecular gas \citep[$L_{\rm bol}\sim10^{47}\rm erg\,s^{-1}$,][also discussion in Sect. \ref{subsec:C2nature}]{Miley_2008,Drouart_2014,Falkendal_2019,Fernandez-Aranda_2024,Fudamoto_2025}. Nevertheless, our finding indicates that the compaion systems are relatively gas-rich.

\subsection{Nature of [\ion{C}{ii}] companions}\label{subsec:C2nature}

In Fig. \ref{fig:4c03}b, we show the $L_{\rm IR}$ versus $L_{[\ion{C}{ii}]}/L_{\rm IR}$ ratio, along with comparison samples: the Spiderweb, a powerful and well-studied $z=2.16$ radio galaxy, which is known to be gas-rich, undergoing a starburst, and residing in a dense environment \citep[e.g.,][]{Miley_2006,Seymour_2012,Gullberg_2016,Emonts_2018,DeBreuck_2022}; $z>5$ radio-loud quasar hosts \citep[five with \textnormal{[\ion{C}{ii}]} detections from][]{Mazzucchelli_2025}; $4<z<6$ star-forming galaxies \citep[ALPINE-CRISTAL, e.g.,][]{Bethermin_2020, Faisst_2020, Mitsuhashi_2024}. Even though five out of eight of our companions only have $L_{\rm IR}$ upper limits, we find that our sample is on the same inversely correlated trend as the comparison samples \citep[see also][]{Diaz-santos_2013}. Our sample has a median $L_{\rm [\ion{C}{ii}]}/L_{\rm IR}$ ratio of $9.4\times10^{-4}$ which may also suggest an impact of bright quasars on C$^{+}$ ions, i.e., excitation to higher energy states \citep[e.g.,][]{Mazzucchelli_2025}. J0121-[\ion{C}{ii}]-B and 4C19-[\ion{C}{ii}] relatively agree with Spiderweb CGM. This indicates that the environment of some of our $z\approx3.5$ radio galaxies is similar to the ones at $z\sim2$. As \citet{DeBreuck_2022} Spiderweb [\ion{C}{ii}] observation was carried out using single-dish data with limited spatial information, it is possible that the cold gas, both of their CGM and AGN components, does also not coincide spatially with the radio AGN host in Spiderweb \citep[see also,][]{Gullberg_2016}.

Taking the zeroth-order estimate of $M_{\rm dyn}$ and assuming $M_{\star}=M_{\rm dyn}/2$ (Sect. \ref{sec:analysis}), we constrain our sample on the $M_{\star}$-SFR plane in Fig. \ref{fig:relation}c \citep[SED based $M_{\star}$ and SFR for ALPINE galaxies,][]{Faisst_2020}. For the \citet{Mazzucchelli_2025} sample, we take their estimated $M_{\rm dyn, dis}$ and assume $M_{\star}=M_{\rm dyn}/2$, consistent with our approach. We calculate their SFR using the $L_{\rm FIR}$ reported by \citet{Mazzucchelli_2025}, in the same way as for our sample. We note that the contribution of jet synchrotron emission to the continuum at frequencies around [\ion{C}{ii}] is negligible, $\lesssim0.1\%$, for HzRGs, i.e., our SFR estimates based on FIR continuum are not affected by non-thermal contribution \citep[see][]{Falkendal_2019}. Synchrotron emission might be a source of contamination for type-1 radio quasars such as the \citet{Mazzucchelli_2025} sample which has less steep radio SED index than HzRGs. We caution that, given most of our measurements are upper limits, the positions of our companions on the SFR-$M_{\star}$ plane are uncertain. If we take the $M_{\star}$ of J0121-[\ion{C}{ii}]-B estimated based on the optical continuum, it may reside above the main sequence \citep[yellow-filled diamond with a gray border, $10^{9.9}\,M_{\odot}$,][]{Wang_companion}. Our findings may indicate that these are starbursting clumps. The non-detection of FIR continuum in five out of eight cases may suggest low SFR. However, we note that the star formation timescale probed by FIR dust continuum ($\sim100\,\rm Myr$) is longer than, for example, measures from H$\alpha$ \citep[$\sim10\,\rm Myr$, e.g.,][]{DeLooze_2014}. Hence, there is a possibility that some of these companions are super bursty events. This may further indicate a positive feedback scenario by the radio jet \citep[][]{Nesvadba_2020}.

A more complicated case is the 4C+03.24 system. The [\ion{C}{ii}] companions around 4C+03.24 is the only example in our sample that shows a morphological connection with the jet. This may resemble the discovery by \citet{Walter_2025} where the jet is interacting with [\ion{C}{ii}] in a type-1 quasar at $z\approx5.8$. Specifically, we see that the [\ion{C}{ii}] clouds around 4C+03.24 are distributed along the radio jet and along the optical continuum (Fig. \ref{fig:4c03}). As discussed in \citet{Wang_companion}, 4C+03.24 system may be a post-major-merger where the cold gas was ram pressure stripped \citep[the optical continuum may be merger debris, e.g.,][]{Bekki_2009}. Hence, another possibility is that these [\ion{C}{ii}] emissions may be shock heated by the jet or the merger \citep[see][for a correlation of shock tracer, warm H$_{2}$ emission, and {[\ion{C}{ii}]}]{Ogle_2012,Appleton_2017,Appleton_2018}. Base on the morphology and distribution of 4C03-[\ion{C}{ii}]-A, [\ion{C}{ii}]-C, and the radio jet, it may be that the jet penetrates through these two clouds which may be evidence of jet-gas interaction.

\subsection{Impact on evolution}

Regardless of the nature of the [\ion{C}{ii}] companions, our detections show that there are gas-rich (or pure gas) systems close to the radio AGN. Similar results are seen in other high-$z$ radio quasars. For example, the Dragonfly galaxy, a $z=1.92$ radio galaxy, has a gas-rich ``wet" major merger near the radio AGN \citep[][]{Emonts_2015,Emonts_2015a,ZhongYuxing_2024}. \citet{Vayner_2021} found that large molecular gas reservoirs are not located at the position of the quasar. In addition, \citet{Lee_2023} reported massive H$_2$ companions around a $z=5.2$ radio galaxy. \citet{Wang_companion} suggested that these candidate mergers can be a trigger of radio AGN activities \citep[e.g.,][]{Chiaberge_2015}. Similarly, \citet{Meyer_2025} argued that the [\ion{C}{ii}]-traced gas in a $z\approx6.6$ quasar may be related to merger stripping. In this Letter, we further propose an evolution scenario of the whole system. The massive hosts of the radio AGN formed their stellar mass fast and depleted most of the gas in the host galaxies \citep[$M_{\star}\sim10^{11}$,][]{DeBreuck_2010,Falkendal_2019}. The gas-rich companions go through minor mergers or interactions with the central galaxies through which they brought gas (or gas being passively stripped) into the central supermassive black holes \citep[SMBHs, e.g.,][]{Lehnert_2016}. The interaction facilitates the dissipation of angular momentum, enabling gas accretion, leading AGN ignition and jet launching. The AGN feedback may later compress the gas and start off-center starbursts. Alternatively, the feedback may be negative such that the jets heat up the gas through shocks leading to the observed [\ion{C}{ii}] \citep[][]{Ogle_2012,Vayner_2025}. This further implies the radio AGN may quench the CGM. The active galaxy, NGC 3079, and its companions may be a local example of this quenching where the ``AGN stripping" is happening, i.e., outflow blows gas out of the companions \citep[e.g.,][]{Irwin_1987,Shafi_2015}.


\subsection{Conclusion}
In this Letter, we presented the analysis of cold gas companion systems traced by [\ion{C}{ii}] emissions at $\lesssim18\,\rm kpc$ around four $z\approx3.5$ radio galaxies observed by ALMA in Band 8 \citep[][]{Wang_companion}. The distribution of [\ion{C}{ii}] and comparison with [\ion{O}{iii}]$\lambda$5007 observed by \textit{JWST}/NIRSpec IFU suggests that the majority of the cold gas clouds are located away from the radio core and do not always follow the fully ionized gas. Guided by the high-resolution [\ion{C}{ii}] data, we find that the [\ion{C}{i}], with $\sim10$ times poorer resolution, is located at the companion position. Our sample has a median $L_{[\rm \ion{C}{ii}]}$ of $1.3\times10^{9}\,L_{\odot}$ and is relatively below the $M_{\rm H_{2}}$-$L_{[\ion{C}{ii}]}$ relation with a low H$_{2}$ mass derived from [\ion{C}{i}]. Compared to the cold gas content at the radio core position, we find that our companions are more gas-rich than the AGN hosts. We propose that these [\ion{C}{ii}] companions may ignite the radio AGN by bringing gas into the SMBHs through merger interactions. The [\ion{C}{ii}] emitters may be starbursting clumps and/or shock heated gas implying the feedback from the radio AGN impacting the CGM of dense HzRG systems. Future analysis of line ratios and kinematics with ionization and shock modeling, jointly with optical lines in NIRSpec IFU \citep[e.g.,][]{Wang_2024,Wang_companion,Vayner_2025} and UV lines in VLT/MUSE \citep[especially \ion{C}{iv}1548,1551 and {\ion{C}{iii}]}1907,1909,][]{Kolwa_2019,wang2023}, will reveal the complex scenarios of feedback on kiloparsecs scales. In addition, \textit{JWST}/MIRI observations of direct warm H$_{2}$ detection will be crucial for shock analysis of these systems.

\begin{acknowledgments}
We thank the anonymous referee for their valuable comments and suggestions, which have improved the quality of this manuscript. We thank Matthieu B\'{e}thermin for the discussion of ALMA [\ion{C}{ii}] data. This work is based in part on observations made with the NASA/ESA/CSA \textit{James Webb} Space Telescope. The data were obtained from the Mikulski Archive for Space Telescopes at the Space Telescope Science Institute, which is operated by the Association of Universities for Research in Astronomy, Inc., under NASA contract NAS 5-03127 for JWST. These observations are associated with program JWST-GO-1970. Support for program JWST-GO-1970 was provided by NASA through a grant from the Space Telescope Science Institute, which is operated by the Association of Universities for Research in Astronomy, Inc., under NASA contract NAS 5-03127. W.W. also acknowledges the grant support from NASA through JWST-GO-3045 and JWST-GO-3950. The work of D.S. was carried out at the Jet Propulsion Laboratory, California Institute of Technology, under a contract with NASA. This publication has received funding from the European Union's Horizon 2020 research and innovation programme under grant agreement No 101004719 (ORP). This paper makes use of the following ALMA data: ADS/JAO.ALMA\#2021.1.00576.S., ADS/JAO.ALMA\#2015.1.00530.S. ALMA is a partnership of ESO (representing its member states), NSF (USA) and NINS (Japan), together with NRC (Canada), MOST and ASIAA (Taiwan), and KASI (Republic of Korea), in cooperation with the Republic of Chile. The Joint ALMA Observatory is operated by ESO, AUI/NRAO and NAOJ.
\end{acknowledgments}

\vspace{5mm}
\facilities{ALMA, \textit{JWST}}

\software{
\texttt{astropy} \citep{Astropy_2013,Astropy_2018}; \texttt{Jupyter notebook} \citep[][]{kluyver2016jupyter}; \texttt{matplotlib} \citep[][]{Hunter_2007}; \texttt{SciPy} \citep[][]{virtanen2020scipy}; \texttt{NumPy} \citep[][]{harris2020numpy}; \texttt{LMFIT} \citep[][]{newville2016}.
          }

\appendix

\section{Molecular gas of 4C+03.24}\label{app:H2_4C03}

\begin{figure}
    \centering
    \includegraphics[width=\columnwidth,clip]{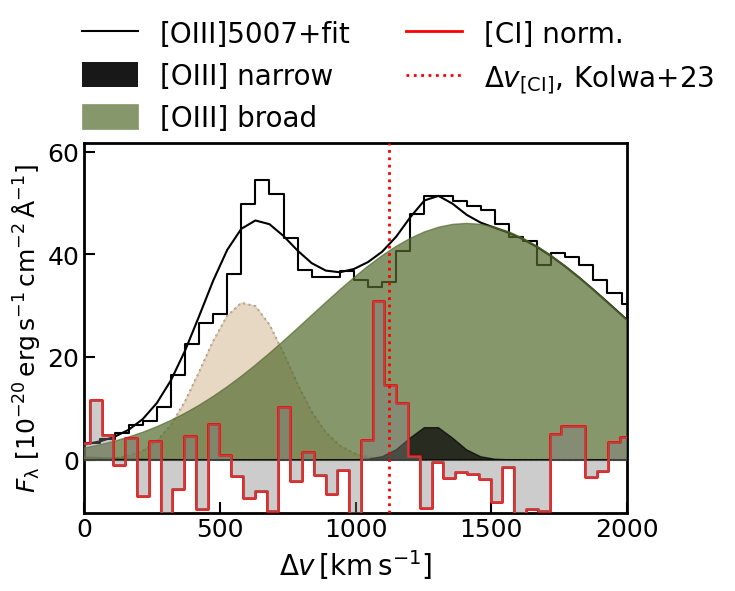}
    \caption{[\ion{O}{iii}]$\lambda$5007 spectrum from $r=0.11\arcsec$ aperture overlaid with normalized [\ion{C}{i}] spectrum at the position of 4C03-[\ion{O}{iii}]. The [\ion{C}{i}] spectrum is extracted at the same position. The black (narrow) and dark green (broad) shaded Gaussians mark the kinematic components reported as a [\ion{O}{iii}] companion in \citet{Wang_companion}. The vertical dotted line marks the $\Delta v$ of the [\ion{C}{i}] detection in \citet{kolwa2023}, $\Delta v_{\rm [\ion{C}{i}],K23}=1123\,\rm km\,s^{-1}$, the same as in Fig. \ref{fig:4c03}b. }
    \label{fig:4C03-OIII}
\end{figure}

\citet{kolwa2023} reported a [\ion{C}{i}] detection of 4C+03.24 at $\Delta v_{\rm [\ion{C}{i}],K23}=1123\,\rm km\,s^{-1}$ with respect to $z_{\rm sys}= 3.5657$ \citep[][]{Nesvadba_2017b}. This emission peak is also seen in our new extraction centered at 4C03-[\ion{C}{ii}]-A, 4C03-[\ion{C}{ii}]-C, and 4C03-[\ion{O}{iii}] (Fig. \ref{fig:4C03-OIII}). We note that  4C03-[\ion{C}{ii}]-A ([\ion{C}{ii}]-C) and 4C03-[\ion{O}{iii}] has a projected separation of $0.5\arcsec$ ($0.9\arcsec$) which is $\sim2-4$ smaller than the Band 3 synthesized beam size. As discussed in Sect. \ref{sec:disc}, we expect to observe [\ion{C}{ii}] emissions when [\ion{C}{i}] is detected (Fig. \ref{fig:tnj0121}b and \ref{fig:j0205_4c19}d). Therefore, this S/N$\simeq2$ and low $M_{\rm H_{2}}\sim6\times10^{9}\,M_{\odot}$ detection may not be real. 

We show the fitted [\ion{O}{iii}]$\lambda$5007 spectrum at the position of 4C03-[\ion{O}{iii}] from \citet{Wang_companion}. The two [\ion{O}{iii}] kinematic components have $\Delta v=1280$ and 1410$\,\rm km\,s^{-1}$, respectively. This is relatively consistent with $\Delta v_{\rm [\ion{C}{i}],K23}$. In \citet{Wang_companion},  we proposed that these emission at projected distance of $2\,$kpc from the radio core but having a large velocity shift may be a merger and/or a second AGN. The detected [\ion{C}{ii}] and optical continuum might be stripped gas clouds and stars, respectively. This [\ion{C}{i}] gas might be leftover in the merger. It might also be the cooled gas induced by compression from merger and/or jet, i.e., the so-called positive feedback \citep[e.g.,][]{Fragile_2004,Fragile_2017,Nesvadba_2020}.

\bibliography{references}{}
\bibliographystyle{aasjournal}

\end{document}